\documentclass[sigconf]{acmart}
\AtBeginDocument{%
  }

\setcopyright{rightsretained}
\copyrightyear{2025}
\acmYear{2025}
\acmConference{SIGGRAPH Posters '25}{August 10-14, 2025}{Vancouver, BC, Canada}\acmBooktitle{Special Interest Group on Computer Graphics and Interactive Techniques Conference Posters (SIGGRAPH Posters '25), August 10-14, 2025}\acmDOI{10.1145/3721250.3743033}
\acmISBN{979-8-4007-1549-5/2025/08}

\usepackage{graphicx}
\usepackage{float}
\usepackage{animate}
\usepackage{subcaption}
\usepackage{wrapfig}

\begin{document}

\title{Two-Stage Sketch-Based Smoke Illustration Generation using Stream Function}

\author{Hengyuan Chang}
\orcid{0009-0005-6628-6560}
\affiliation{
  \institution{Japan Advanced Institute of Science and Technology}
  \city{Nomi}
  \state{Ishikawa}
  \country{Japan}
}

\author{Xiaoxuan Xie}
\orcid{0009-0005-3515-9091}
\affiliation{
  \institution{Japan Advanced Institute of Science and Technology}
  \city{Nomi}
  \state{Ishikawa}
  \country{Japan}
}

\author{Syuhei Sato}
\orcid{0009-0007-9270-7335}
\affiliation{
  \institution{Hosei University}
  \city{Tokyo}
  \country{Japan}
}

\author{Haoran Xie}
\orcid{0000-0002-6926-3082}
\affiliation{
  \institution{Japan Advanced Institute of Science and Technology}
  \city{Nomi}
  \state{Ishikawa}
  \country{Japan}
}

\renewcommand{\shortauthors}{Chang et al.}


\begin{teaserfigure}
\centering
  \includegraphics[width=\linewidth]{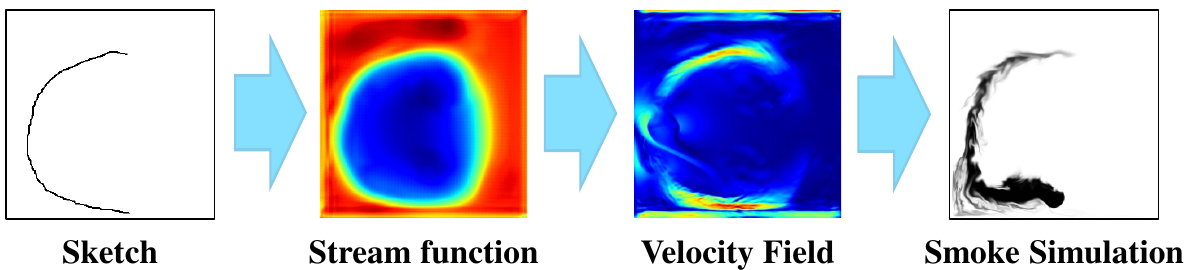}
  \caption{This work first generates the stream function from the user sketch, then obtains the velocity field using the diffusion-based generators. Finally, the generated velocity field can be used as the guidance force to drive smoke simulations. }
  \label{fig:teaser}
\end{teaserfigure}


\maketitle

\section{Introduction}
While smoke effects impart dynamic expressiveness to illustrations, conventional smoke illustration creation usually relies on physics-based fluid simulations, as implemented in Blender and Unreal Engine. However, the operational complexity may impede usage by common users.
Hand-drawn sketches offer a concise structural depiction with lines and curves, serving as a freeform approach for both artists and amateurs. The sketch-based smoke illustration pipelines can simplify the creation process and enhance accessibility in multimedia creation.
Generative models have democratized illustration design, enabling amateurs to create illustrations. The previous smoke illustration synthesis approaches are primarily based on conditional GANs~\cite{xie2024dualsmoke}, which exhibit training instability and mode collapse under limited data regimes. Existing diffusion-based methods relatively rely on text and pixel features, with limited incorporation of physical priors~\cite{chang2025diffsmoke}. 

To solve the above issues, we propose a two-stage sketch-based smoke illustration generation framework using stream function and latent diffusion models (LDM)~\cite{rombach2022high} (Figure~\ref{fig:teaser}). The user sketch is used to guide the generation of the stream function, which serves as the control condition for the velocity field generator. The generated velocity field can be used to guide the smoke simulation to align with the intended flow.
We adopt streamlines to encode global flow dynamics as sketch guidance during training. The stream function constitutes the intermediate representation that captures continuous variation and rotational flow details absent from sketches.

\begin{figure*}[t]
  \includegraphics[width=0.99\linewidth]{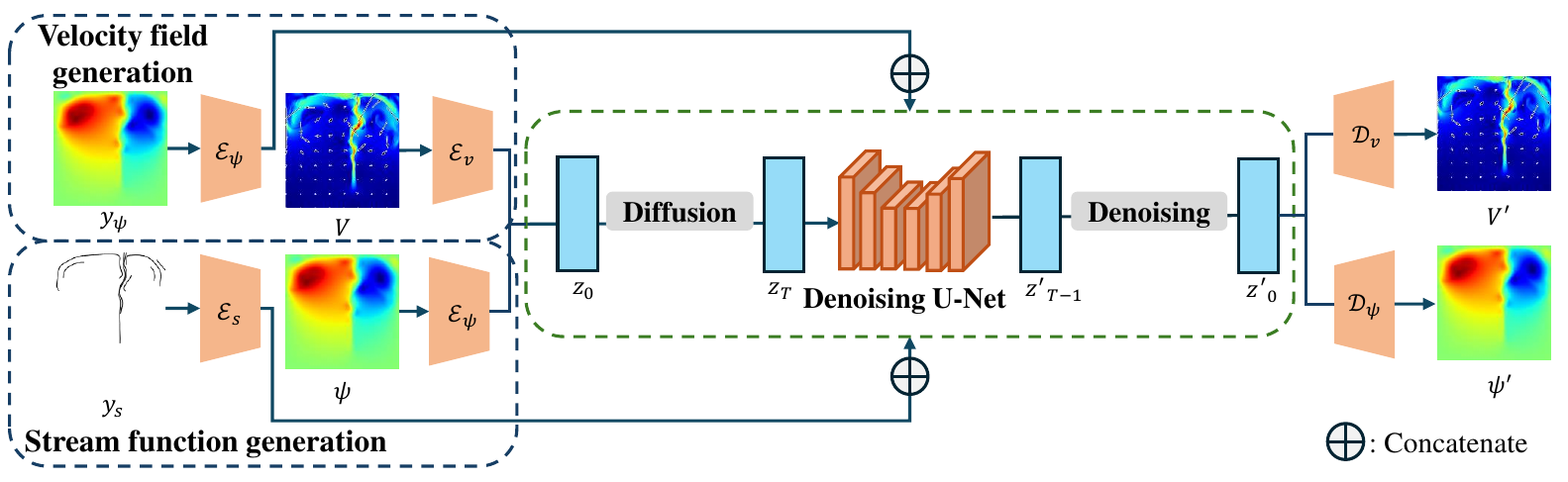}
  \caption{The workflow of our proposed sketch-based smoke illustration framework using stream function.}
  \label{fig:framework}
\end{figure*}

\section{Methods}
Figure \ref{fig:framework} illustrates the workflow of our proposed framework for fluid video generation guided by hand-drawn sketches. 

\noindent
\textbf{\textit{Smoke Simulation}}. We implement incompressible 2D smoke simulation scenarios for training data collection. The velocity $\textbf{u}$ at each time step $t$ is evolved by the inviscid Navier-Stokes equations and boundary conditions given as follows: 
\begin{align}
    \frac{\partial \textbf{u}}{\partial t} = -\textbf{u}\cdot\nabla\textbf{u}-\frac{1}{\rho}\nabla p &+ {\textbf{f}_e}\,, \; \nabla \cdot \textbf{u} = 0\\
    \textbf{u} \cdot \textbf{n} &= 0
\end{align}
where $\mathbf{u}$ is 2D velocity field, $p$ is pressure, $\rho$ is density, external force $\textbf{f}_e$ is composed of horizontal and vertical constant forces, and $\mathbf{n}$ is normal vector. The semi-Lagrangian method is used for time integration.

\noindent
\textbf{\textit{Stream Function}}. The stream function represents rotational and continuous dynamics of the velocity field. Positive and negative values indicate opposite directions of rotation. We extract the stream function via Helmholtz-Hodge decomposition from a 2D velocity field~\cite{sato2021stream}.
The decomposition and stream function calculation are given as follows,
\begin{align}
    \mathbf{U} = \nabla P + \nabla \times \mathbf{\psi} + \mathbf{H}\\
    \nabla^2 \psi = -\nabla \times \mathbf{U}
\end{align}
where $\mathbf{U}$ represents a 2D velocity field. The stream function $\psi$ is a 2D scalar field, and the scalar field $P$ is calculated from the divergence of the velocity field. 
In a connected domain with no flux boundaries, $P$ and $\mathbf{H}$ vanish due to the incompressibility, with only the divergence-free component remaining. We solve the above equation using a conjugate gradient solver.

\noindent
\textbf{\textit{Streamline}}. Streamlines can intuitively depict the instantaneous direction and structure of the velocity field. Hence, streamlines are employed as sketch data for training in this study. 
We seeded particles at grid centers and integrated the trajectories using the Runge-Kutta method. We select the streamlines originating from 512 particles with the highest velocity magnitude. 

\noindent
\textbf{\textit{Two-Stage Latent Diffusion Model}}.
Our training framework adopted a two-stage generation model as illustrated in Figure~\ref{fig:framework}.
Sketch input guides the stream function generation. Residual noise in the synthesized stream function propagates through the curl operator into the derived velocity field. Hence, we generate the velocity field with a diffusion model to attenuate the effect of noise.

\section{Results}
We conducted a comparative analysis with DualSmoke\cite{xie2024dualsmoke} and DiffSmoke\cite{chang2025diffsmoke}. 
The MSE(mean squared error) loss is selected as the training loss. 
The quantitative evaluation results are given in Table~\ref {tab:1}.
\begin{table}[htb]
\centering
\scalebox{0.8}{
\begin{tabular}{cccccc}
\toprule
Methods &DualSmoke & DiffSmoke &Ours \\
\midrule
MSE Loss (First stage)&0.050&0.031&\textbf{0.005}\\
\midrule
MSE Loss (Second stage)&0.572&0.052&\textbf{0.035}\\
\bottomrule
\end{tabular}
}
\caption{MSE loss comparison. First stage is stream function generation, second stage is for velocity field generation. }
\label{tab:1}
\end{table}

\begin{figure}[t]
    \centering
    \includegraphics[width=0.99\linewidth]{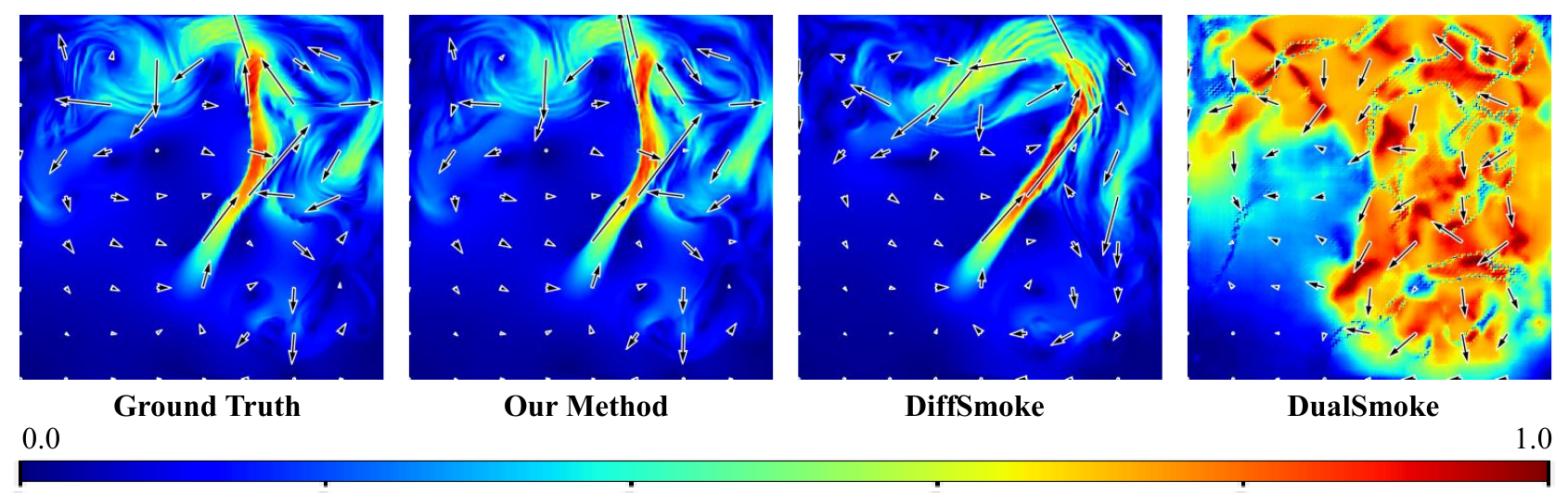}
    \caption{The qualitative evaluation result samples. The visualized velocity field is normalized into [0, 1] range. }
    \label{eval}
\end{figure}

\begin{figure}[ht]
    \centering
    \includegraphics[width=0.99\linewidth]{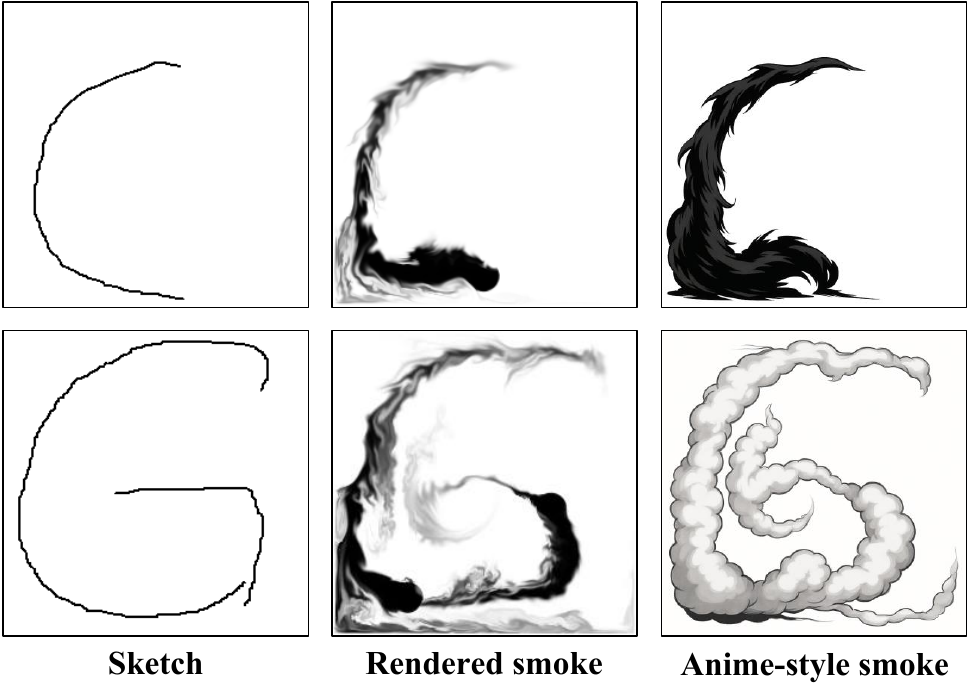}
    \caption{The anime style transferred smoke samples. }
    \label{anime}
\end{figure}

DualSmoke model underperformed on our collected dataset. Although the fluid spatial distribution was successfully captured, the field magnitudes were noisy.  DiffSmoke recovered the global flow topology, but the local detail reconstruction remains suboptimal, exhibiting inaccuracies in both direction and magnitude. 
The qualitative result is presented in Figure ~\ref{eval}.
The generated velocity field is used as a guidance force field in the smoke simulation~\cite{xie2024dualsmoke}. In addition, we conducted an anime-style transfer LLM interface after the simulation module for the generation of anime-style smoke illustrations. The results are given in Figure~\ref{anime}.

\section{Conclusion}
In this work, we proposed a sketch-based two-stage smoke illustration generation framework using stream function, which outperforms existing smoke illustration generation studies in both quantitative metrics and qualitative evaluations.

\begin{acks}
This work is funded by New Energy and Industrial Technology Development Organization (NEDO) JPNP20017, JSPS KAKENHI Grant Numbers 23K18514 and 25K00154.
\end{acks}
\bibliographystyle{ACM-Reference-Format}
\bibliography{sample-base}


\end{document}